\documentclass[prl,aps,nofootinbib,showkeys,showpacs,twocolumn]{revtex4-1}

\usepackage[T1]{fontenc} 
\usepackage{epsfig}
\usepackage{graphicx}
\usepackage{amssymb}
\usepackage{ulem}
\usepackage{amsmath}
\usepackage{color}
\definecolor{darkgreen}{rgb}{0.2,0.5, 0.2}

\usepackage{dcolumn,booktabs}
\newcolumntype{d}[1]{D{.}{.}{#1}}

\definecolor{mbscolor}{rgb}{0.60, 0.0, 0.65}

\usepackage{dsfont}

\renewcommand\sout{\bgroup \color{red} \ULdepth=-.5ex \ULset}

\begin{document}
\title{Constraints on the Symmetry Energy from PREX-II in the Multimessenger Era}

\author{Tong-Gang Yue$^{1}$}
\author{Lie-Wen Chen$^{1}$}
\email{Corresponding Author: lwchen$@$sjtu.edu.cn}
\author{Zhen Zhang$^{2}$}
\author{Ying Zhou$^{3}$}
\address{$^{1}$School of Physics and Astronomy, Shanghai Key Laboratory for
Particle Physics and Cosmology, and Key Laboratory for Particle Astrophysics and Cosmology (MOE),
Shanghai Jiao Tong University, Shanghai 200240, China}
\address{$^{2}$Sino-French Institute of Nuclear Engineering and Technology,
Sun Yat-sen University, Zhuhai 519082, China}
\address{$^{3}$Quantum Machine Learning Laboratory, Shadow Creator Inc., Shanghai 201208, China}

\date{\today}

\begin{abstract}
The neutron skin thickness $\Delta r_{\rm{np}}$ of heavy nuclei is essentially determined by the symmetry energy density slope $L({\rho })$ at $\rho_c = 0.11~{\rm {fm}^{-3}}\approx 2/3\rho_0$ ($\rho_0$ is nuclear saturation density), roughly corresponding to the average density of finite nuclei.
The PREX collaboration
recently reported a model-independent extraction of $\Delta r^{208}_{\rm{np}} = 0.283 \pm 0.071$~fm for the $\Delta r_{\rm{np}}$ of $^{208}$Pb, suggesting a rather stiff symmetry energy $E_{\rm{sym}}({\rho })$ with $L({\rho_c }) \ge 52$~MeV.
We show that the $E_{\rm{sym}}({\rho })$ cannot be too stiff and $L({\rho_c }) \le 73$~MeV is necessary to be compatible with
(1) the ground-state properties and giant monopole resonances of finite nuclei, (2) the constraints on the equation of state of symmetric nuclear matter at suprasaturation densities from flow data in heavy-ion collisions, (3) the largest neutron star (NS) mass reported so far for PSR J0740+6620, (4) the NS tidal deformability extracted from gravitational wave signal GW170817 and (5) the mass-radius of PSR J0030+045 measured simultaneously by NICER.
This allows us to obtain $52 \le L({\rho_c }) \le 73$~MeV and $0.212 \le \Delta r^{208}_{\rm{np}} \le 0.271$~fm, and further $E_{\rm{sym}}({\rho_0 }) = 34.3 \pm 1.7$~MeV, $L({\rho_0 }) = 83.1 \pm 24.7$~MeV, and $E_{\rm{sym}}({2\rho_0 }) = 62.8 \pm 15.9$~MeV. A number of critical implications on nuclear physics and astrophysics are discussed.
\end{abstract}

\maketitle

\textit{Introduction.}---
The Lead Radius Experiment~(PREX) collaboration recently reported a model-independent extraction of $\Delta r^{208}_{\rm{np}} = 0.283 \pm 0.071$~fm~\cite{PREX-II} for the neutron skin thickness (the difference between the rms radii of the neutron and proton distributions, $\Delta r_{\rm{np}} \equiv r_n - r_p$) of $^{208}$Pb by combining the original PREX result~\cite{Abr12} with the new PREX-II measurement~\cite{PREX-II}. This updated result (hereafter referred to as simply ``PREX-II'') reaches a precision close to $1\%$ for $r_n$, much more precise than the original $\Delta r^{208}_{\rm{np}} = 0.33^{+0.16}_{-0.18}$~fm~\cite{Abr12}.
In PREX, the neutron density distribution in $^{208}$Pb is determined by measuring the parity-violating electroweak asymmetry in the elastic scattering of polarized electrons off $^{208}$Pb and thus is free from the strong interaction uncertainties.
Since the proton is charged and its distributions are well determined, the $\Delta r^{208}_{\rm{np}} = 0.283 \pm 0.071$~fm may represent the cleanest and most accurate $\Delta r^{208}_{\rm{np}}$ so far although the more precise measurement has been planned at MESA~\cite{Bec18}. The coherent elastic neutrino-nucleus scattering~\cite{Aki17} provides another clean way to extract the $\Delta r_{\rm{np}}$, but the current uncertainty is too large~\cite{Cad18,HuangXR19}.
The $0.283 \pm 0.071$~fm means a rather thick $\Delta r^{208}_{\rm{np}}$, significantly larger than those extracted from other approaches that suffer from the uncertainties of the strong interaction (see, e.g., Ref.~\cite{Thi19} for a recent review).

Besides its fundamental importance for nuclear structure, the $\Delta r_{\rm{np}}$ has been identified as an ideal probe on the symmetry energy $E_{\rm{sym}}({\rho })$ --- a key but poorly-known quantity that encodes the isospin dependence of nuclear matter equation of state (EOS) and plays a critical role in many issues in nuclear physics and astrophysics~\cite{Ste05,LCK08,Gan15,ZhangNB19,Oze16,Bal16}.
Indeed, it has been established~\cite{Bro00,Fur02,ChenLW05PRC,Cen09,ChenLW10,Roc11} that the $\Delta r_{\rm{np}}$ exhibits a strong positive linear correlation with the symmetry energy density slope $L({\rho })$ at nuclear saturation density $\rho_0 \approx 0.16$~fm$^{-3}$, i.e., $L \equiv L({\rho_0 })$.
An even stronger correlation is found between the $\Delta r_{\rm{np}}$ of heavy nuclei and the $L({\rho })$ at a subsaturation cross density $\rho_c = 0.11~{\rm {fm}^{-3}}\approx 2/3\rho_0$~\cite{ZhangZ13}, roughly corresponding to the average density of finite nuclei, i.e., $L_c \equiv L({\rho_c })$.
Furthermore, the $L(\rho)$ around $\rho_0$ strongly influences the mass-radius~(M-R) relation and tidal deformability of neutron stars~(NSs), and thus provides a unique bridge between atomic nuclei and NSs~\cite{Tod05,Fat18,ZhouY19PRD,ZhouY19ApJ}.

The large value of $\Delta r^{208}_{\rm{np}} = 0.283 \pm 0.071$~fm
suggests a very stiff $E_{\rm{sym}}({\rho })$ (a large $L({\rho })$) around $\rho_0 $, which generally leads to a very large NS radius and tidal deformability.
However,
an upper limit of $\Lambda_{1.4} \le 580$~\cite{Abb18NSMerger} for the dimensionless tidal deformability of $1.4M_\odot$ NS has been obtained from the gravitational wave signal GW170817, which requires a relatively softer $E_{\rm{sym}}({\rho })$.
In addition,
the heaviest NS with mass $2.14^{+0.10}_{-0.09}M_\odot$ for PSR J0740+6620~\cite{Cro19Mmax} also strongly limits the $E_{\rm{sym}}({\rho })$~\cite{ZhouY19ApJ}, especially under the constraints on the EOS of symmetric nuclear matter~(SNM) at suprasaturation densities from flow data in
heavy-ion collisions~(HIC)~\cite{Dan02}, which is relatively soft and strongly restricts the NS maximum mass $M_{\rm max}$~\cite{ZhouY19PRD,ZhouY19ApJ,CaiBJ17}.
Furthermore,
two independent simultaneous M-R determinations from NICER~\cite{Ril19NICER,Mil19NICER} for PSR J0030+0451 with mass around $1.4M_\odot$ has been obtained, further constraining the $E_{\rm{sym}}({\rho })$.
Given the rich multimessenger data,
it is extremely important to develop a unified framework that can simultaneously describe the finite nuclei and NSs which involve a very large density range.
Actually,
serious tension between $\Delta r^{208}_{\rm{np}} = 0.283 \pm 0.071$~fm and the limits from GW170817 and NICER has been observed in a covariant density functional study~\cite{Ree20}.

In this work,
within a single unified framework of the extended Skyrme-Hartree-Fock (eSHF) model~\cite{Cha09,ZhangZ16} which includes momentum dependence of effective many-body forces,
we find the $L_c$ cannot be larger than $73$~MeV under the constraints from GW170817, NICER, the NS mass $2.14^{+0.10}_{-0.09}M_\odot$, flow data in heavy-ion collisions, and the data of ground-state properties and giant monopole resonances~(GMR) of finite nuclei.
Our findings produce an upper limit of $\Delta r^{208}_{\rm{np}} \le 0.271$~fm, and this together with the $\Delta r^{208}_{\rm{np}} = 0.283 \pm 0.071$~fm lead to stringent constraints of $0.212 \le \Delta r^{208}_{\rm{np}} \le 0.271$~fm and correspondingly $52 \le L_c \le 73$~MeV,
which have a number of critical implications in nuclear physics and astrophysics.

\textit{Model and method.}---
The EOS of nuclear matter at density $\rho=\rho_{\rm n}+\rho_{\rm p}$ and isospin asymmetry $\delta=(\rho_{\rm n}-\rho_{\rm p})/\rho$
with $\rho_{\rm n}$($\rho_{\rm p}$) denoting the neutron(proton) density, defined by the binding energy per nucleon,
can be expressed as
\begin{equation}\label{EOS}
E(\rho,\delta)=E_0(\rho)+E_{\rm sym}(\rho)\delta^2+{\cal O}(\delta^4),
\end{equation}
where $E_0(\rho)=E(\rho,\delta=0)$ is SNM EOS and $E_{\rm{sym}}(\rho) = \left.\frac{1}{2!}\frac{\partial^{2}E(\rho,\delta)}{\partial\delta^{2}}\right|_{\delta=0}$
is the symmetry energy.
At $\rho_0$, the $E_0(\rho)$ can be expanded in
$\chi=(\rho-\rho_0)/(3\rho_0)$ as
$E_0(\rho)=E_0(\rho_0)+\frac{1}{2!}K_0\chi^2+\frac{1}{3!}J_0\chi^3+{\cal O}(\chi^4)$,
in terms of incompressibility $K_0$ and skewness $J_0$.
The $E_{\rm sym}(\rho)$ can be expanded at a reference density
$\rho_r$ in terms of the slope parameter $L(\rho_r)$ and the curvature
parameter $K_{\rm{sym}}(\rho_r)$ as
$E_{\rm{sym}}(\rho) = E_{\rm{sym}}(\rho_r) + L(\rho_r) \chi_r + \frac{1}{2!}K_{\rm{sym}}(\rho_r)\chi_r^2+\mathcal{O}(\chi_r^3)$,
with $\chi_r=(\rho-\rho_r)/(3\rho_r)$. Setting $\rho_r = \rho_0$ leads to the conventional $L \equiv L(\rho_0)$ and
$K_{\rm{sym}} \equiv K_{\rm{sym}}(\rho_0)$.

Within the eSHF model~\cite{Cha09,ZhangZ16} which includes $13$ Skyrme interaction parameters $\alpha$,
$t_0\sim t_5$, $x_0\sim x_5$ and the spin-orbit coupling constant $W_0$, we have
\begin{eqnarray}
\label{Eq:E0}
E_0(\rho)&=&\frac{3\hbar^2}{10m}k_{F}^2+\frac{3}{8}t_0\rho + \frac{3}{80}[3t_1+t_2(4x_2+5)]{\rho}k_{F}^2 \notag\\
&+& \frac{1}{16}t_3\rho^{\alpha+1} + \frac{3}{80}[3t_4+t_5(4x_5+5)]{\rho^2}k_{F}^2,
\end{eqnarray}
and
\begin{eqnarray}
\label{Eq:Esym}
E_{\mathrm{sym}}(\rho)
&=&\frac{\hbar^2}{6m}k_{F}^2-\frac{1}{8}t_0(2x_0+1)\rho
-\frac{1}{48}t_3(2x_3+1)\rho^{\alpha+1}\notag\\
&-&\frac{1}{24}[3t_1x_1 -t_2(4+5x_2)]\rho k_{F}^2\notag\\
&-&\frac{1}{24}[3t_4x_4 -t_5(4+5x_5)]\rho^2 k_{F}^2,
\end{eqnarray}
where $m$ is the nucleon rest mass and $k_F =(3\pi ^2/2\rho)^{1/3}$ is the Fermi momentum.
The last term in Eqs.~(\ref{Eq:E0}) and (\ref{Eq:Esym})
is from the momentum dependence of three-body forces which is not considered in the standard SHF model (see, e.g., Ref.~\cite{Cha97}).
The eSHF provides a successful framework to describe simultaneously nuclear matter, finite nuclei,
and NSs~\cite{ZhangZ16}.
The $13$ Skyrme parameters $\alpha$,
$t_0\sim t_5$, $x_0\sim x_5$ can be expressed explicitly in terms of
the following $13$ macroscopic quantities (pseudo-parameters)~\cite{ZhangZ16}:
$\rho_0$, $E_{0}(\rho_0)$, $K_0$, $J_0$,
$E_{\rm sym}(\rho_r)$, $L(\rho_r)$, $K_{\rm sym}(\rho_r)$,
the isoscalar effective mass $m_{s,0}^{\ast}$,
the isovector effective mass $m_{v,0}^{\ast}$,
the gradient coefficient $G_S$,
and the symmetry-gradient coefficient $G_V$,
the cross gradient coefficient $G_{SV}$,
and the Landau parameter $G_0'$ of SNM in the spin-isospin channel.
Instead of directly using the $13$ Skyrme parameters,
we use here the $13$ macroscopic model parameters in the eSHF calculations for nuclear matter,
finite nuclei and NSs~\cite{ZhangZ16}.

For NSs, we consider the conventional NS model, i.e.,
the NS contains core, inner crust and outer crust with the core
including only neutrons, protons, electrons and possible muons ($npe\mu$).
For the details, one is referred to Refs.~\cite{ZhangZ16,ZhouY19PRD,ZhouY19ApJ}.
We would like to emphasize that
in the following NS calculations,
the core EOS is obtained from full eSHF calculations with model parameters constrained 
by properties of both finite nuclei and NSs as well as HIC flow data, and
the core-crust transition density $\rho_{\rm t}$ is determined self-consistently
by a dynamical approach~\cite{XuJ09ApJ}. In addition, all
the constructed eSHF parameter sets used in the following NS
calculations are required to satisfy the causality condition.

\textit{Result and discussion.}---
For the $13$ macroscopic model parameters in eSHF, we fix $E_{\rm sym}(\rho_c) = 26.65$~MeV since it has been obtained with high precision
by analyzing the binding energy difference of heavy isotope pairs~\cite{ZhangZ13}.
Furthermore, the $L_c$ essentially determines the $\Delta r_{\rm{np}}$ of heavy nuclei~\cite{ZhangZ13}, while
the higher-order parameters $J_0$ and $K_{\rm sym}$ only weakly affect
the properties of finite nuclei but are critical for NS properties~\cite{ZhouY19PRD,ZhouY19ApJ}.
To explore the $\Delta r_{\rm{np}}$ and NSs,
therefore, our strategy is to search for the parameter space of $L_c$, $J_0$ and $K_{\rm sym}$ under the constraints on the SNM EOS from flow data as well as the limits from GW170817 and NS observations, while with the other $10$ parameters
($\rho_0$, $E_{0}(\rho_0)$, $K_0$, $m_{s,0}^{\ast}$, $m_{v,0}^{\ast}$, $G_S$, $G_V$, $G_{SV}$, $G_0'$, and $W_0$) being obtained by fitting the nuclear data on the binding energies, charge rms radii, GMR energies, and spin-orbit energy level splittings (see Refs.~\cite{ZhangZ16,ZhouY19PRD,ZhouY19ApJ} for details) to guarantee that the eSHF can successfully describe nuclear properties (the relative deviations of charge radii and total binding energies for medium and heavy nuclei from data are less than $0.5\%$).
From the obtained $L_c$, $J_0$ and $K_{\rm sym}$,
one can extract information on EOS, $\Delta r_{\rm{np}}$, and NSs.

\begin{figure}[tbp]
\centering
\includegraphics[width=1.0\linewidth]{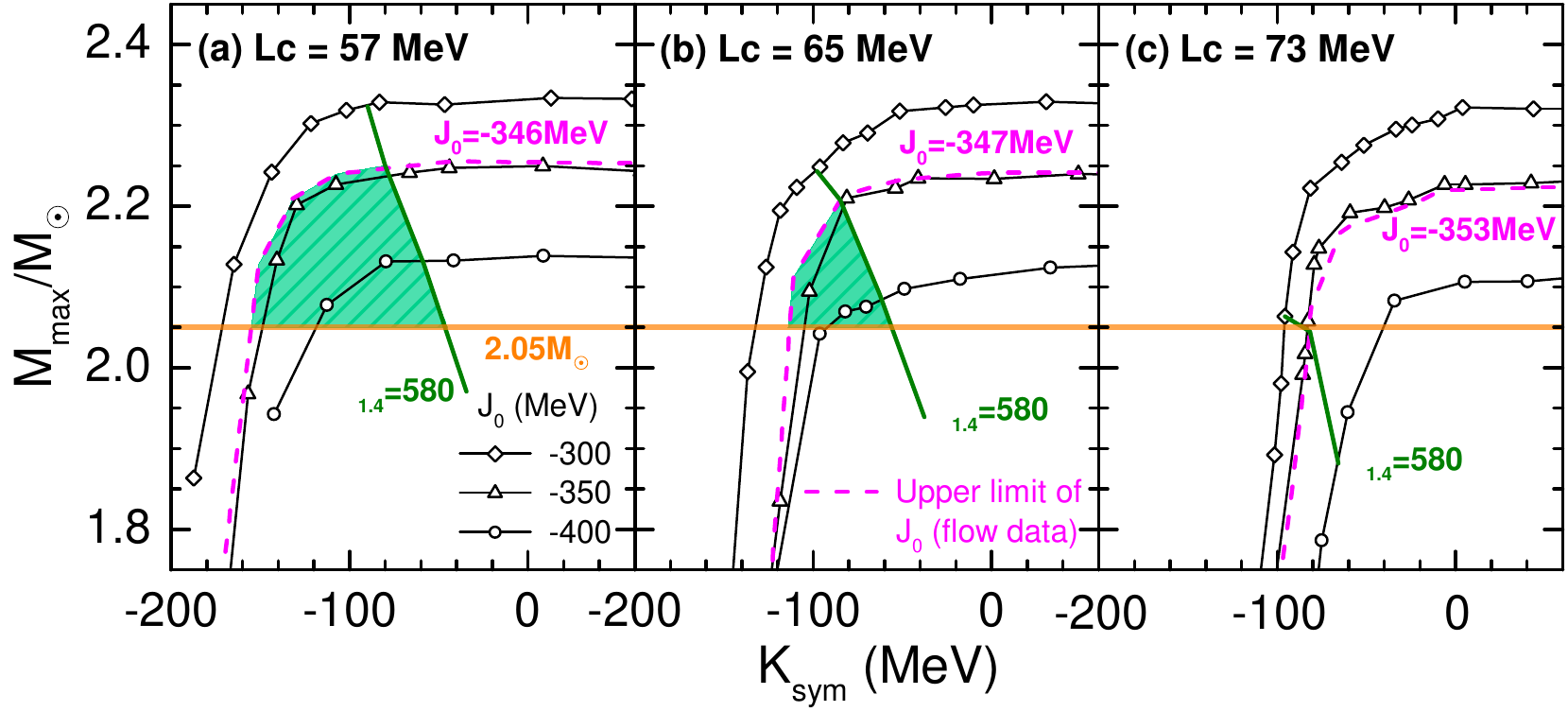}
\caption{NS maximum mass $M_{\rm max}$ vs $K_{\rm sym}$
within the eSHF model in a series of extended Skyrme interactions with
$J_0$ and $K_{\rm sym}$ fixed at various values
for $L_c = 57$~MeV~(a), $65$~MeV~(b) and $73$~MeV~(c), respectively.
The shadowed regions indicate the allowed parameter space.
See the text for details.}
\label{Fig_MmaxKsym}
\end{figure}

A larger $L_c$ generally leads to a larger $\Delta r^{208}_{\rm{np}}$ and correspondingly a larger $\Lambda_{1.4}$. For fixed $L_c$ and $J_0$, reducing the $K_{\rm sym}$ can effectively reduce the $\Lambda_{1.4}$ but also reduces the NS maximum mass $M_{\rm max}$~\cite{ZhouY19PRD,ZhouY19ApJ}. Furthermore, increasing $J_0$ can enhance significantly the $M_{\rm max}$ but the $J_0$ cannot be too large~\cite{ZhouY19PRD,ZhouY19ApJ,CaiBJ17} due to relatively soft SNM EOS constrained by the flow data.
Using the limit of $\Lambda_{1.4} \le 580$ from GW170817, $M_{\rm{max}} \ge 2.05M_\odot$ from PSR J0740+6620, and the flow data constraint on SNM EOS, one thus expects there should exist an upper limit for $L_c$ (also for $J_0$ and $K_{\rm sym}$).
Figs.~\ref{Fig_MmaxKsym}~(a), (b), and (c) show the $M_{\rm max}$ vs $K_{\rm sym}$ at various $J_0$ with $L_c = 57, 65$ and $73$~MeV, respectively. The shadowed regions represent the allowed parameter space of $J_0$ and $K_{\rm sym}$, which all satisfy the limits of $\Lambda_{1.4} \le 580$, $M_{\rm{max}} \ge 2.05M_\odot$, and the flow data constraint. We note that the allowed parameter space agrees with $\Delta r^{208}_{\rm{np}} \ge 0.212$~fm. As expected, one sees from Fig.~\ref{Fig_MmaxKsym} that the allowed parameter space becomes smaller and smaller with increasing $L_c$ (see also Ref.~\cite{ZhouY19ApJ}), and it is essentially reduced to a point at $L_c = 73$~MeV with $K_{\rm sym} = -82$~MeV and $J_0 = -353$~MeV as shown in Fig.~\ref{Fig_MmaxKsym}~(c) (the corresponding parameter set is denoted as ``Lc73''). Therefore, our results indicate the $L_c$ has an upper limit of $L_c \le 73$~MeV.

We note that the eSHF with Lc73 predicts $\Delta r^{208}_{\rm{np}} = 0.268$~fm, which is consistent with the $\Delta r^{208}_{\rm{np}} = 0.283 \pm 0.071$~fm from PREX-II.
On the other hand,
a smaller $L_c$ will lead to a smaller $\Delta r^{208}_{\rm{np}}$ and thus may violate the constraint $\Delta r^{208}_{\rm{np}} = 0.283 \pm 0.071$~fm. Therefore, the lower limit of $\Delta r^{208}_{\rm{np}} = 0.212$~fm from PREX-II can set a lower limit of $L_c$.
To obtain a quantitative relation between $L_c$ and $\Delta r^{208}_{\rm{np}}$, we construct a series of parameter sets with $L_c$ from $30$ to $90$~MeV in a step of $5$~MeV.
For the $9$ parameter sets of
$L_c = 30 \sim 70$~MeV, the $J_0$ and $K_{\rm sym}$ are obtained by requiring them to reach the largest NS mass under the constraints of $\Lambda_{1.4} \le 580$ and $M_{\rm{max}} \ge 2.05M_\odot$ as well as the flow data constraints on SNM EOS.
For the $4$ parameter sets of
$L_c = 75 \sim 90$~MeV, the constraint of $\Lambda_{1.4} \le 580$ is turned off since it cannot be satisfied for $L_c \ge 75$~MeV under the constraints of $M_{\rm{max}} \ge 2.05M_\odot$ and the flow data as discussed earlier.

\begin{figure}[tbp]
\centering
\includegraphics[width=0.9\linewidth]{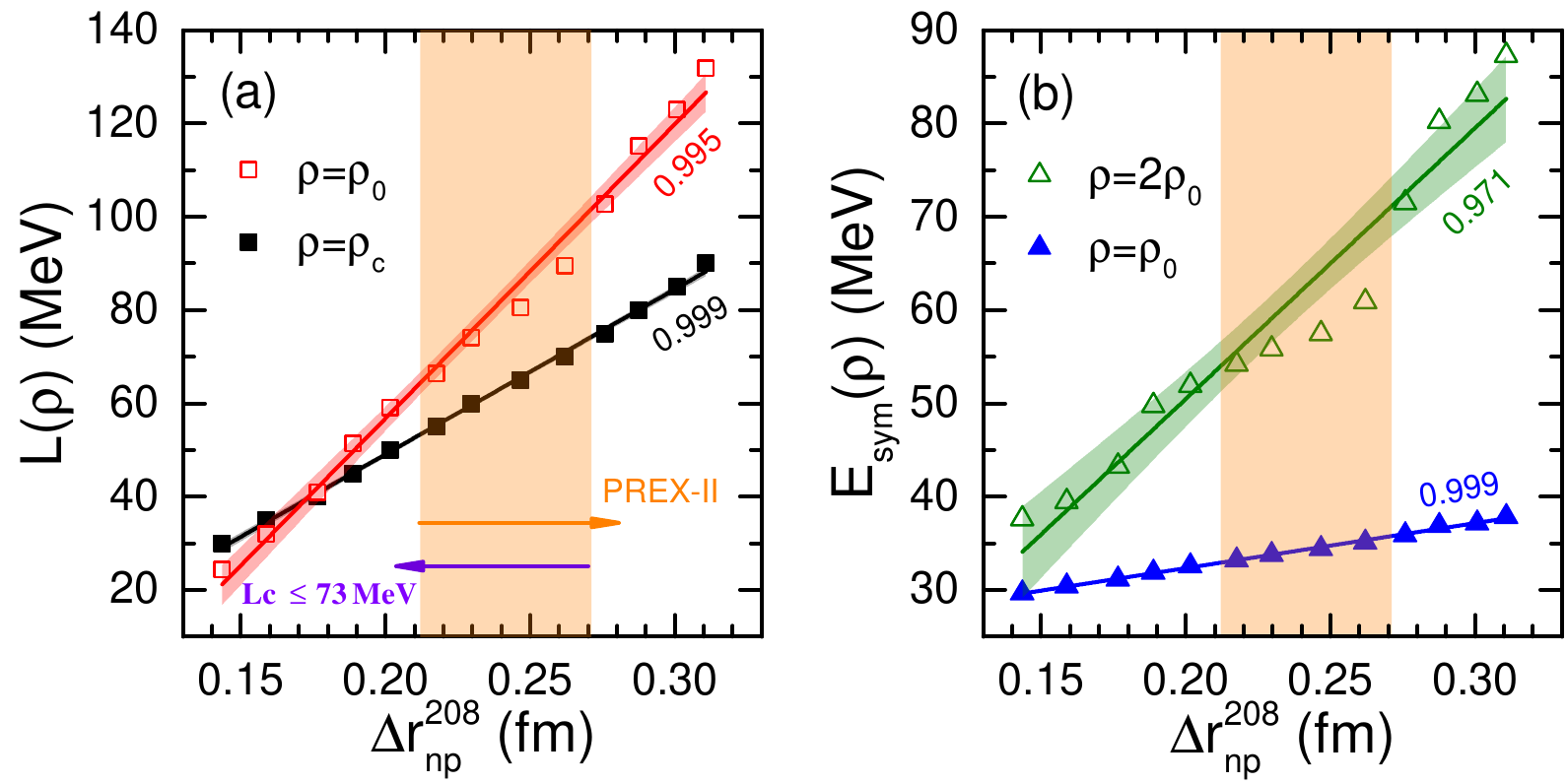}
\caption{The correlation of $\Delta r^{208}_{\rm{np}}$ with $L_c$ and $L$~(a) as well as $E_{\rm sym}(\rho_0)$ and $E_{\rm sym}(2\rho_0)$~(b) in the eSHF model. The limit of $0.212 \le \Delta r^{208}_{\rm{np}} \le 0.271~{\rm fm}$ from PREX-II~\cite{PREX-II} and $L_c \le 73$~MeV obtained in this work is indicated by the orange band.}
\label{Fig_EsymNSkin}
\end{figure}

Using the constructed
$13$ parameter sets, we plot the $L_c$ vs $\Delta r^{208}_{\rm{np}}$ in Fig.~\ref{Fig_EsymNSkin}~(a).
As expected, the $L_c$ displays a very strong positive linear correlation with $\Delta r^{208}_{\rm{np}}$, i.e.,
\begin{equation}\label{Eq:Lc-NSkin}
L_c = (-21.65\pm 1.02) + (353.78\pm 4.32)\Delta r^{208}_{\rm{np}},
\end{equation}
or
\begin{equation}\label{Eq:NSkin-Lc}
\Delta r^{208}_{\rm{np}} = (0.0615\pm 0.0022) + (0.00282\pm 0.0000344)L_c,
\end{equation}
where the units of $L_c$ and $\Delta r^{208}_{\rm{np}} $ are MeV and fm, respectively.
Using Eq.~(\ref{Eq:Lc-NSkin}), one obtains a lower limit of $L_c = 52.0$~MeV with $\Delta r^{208}_{\rm{np}} = 0.212$~fm.
It is interesting to note that using $L_c = 73$~MeV in Eq.~(\ref{Eq:NSkin-Lc}) leads to an upper limit of $\Delta r^{208}_{\rm{np}} = 0.271$~fm, nicely consistent with the eSHF prediction with Lc73. We thus conclude $0.212~{\rm fm} \le \Delta r^{208}_{\rm{np}} \le 0.271~{\rm fm}$ and correspondingly $52~{\rm MeV} \le L_c \le 73~{\rm MeV}$.

To see the implications of the constraint $0.212~{\rm fm} \le \Delta r^{208}_{\rm{np}} \le 0.271~{\rm fm}$ on the symmetry energy, we also show in Fig.~\ref{Fig_EsymNSkin} the $\Delta r^{208}_{\rm{np}}$ vs $L$, $E_{\rm sym}(\rho_0)$ and $E_{\rm sym}(2\rho_0)$, which all display strong linear correlations. To understand these correlations, it is instructive to write down
\begin{equation}\label{Eq:Lr}
L({\rho_r}) \approx L \rho_r/\rho_0 + K_{\rm{sym}} \rho_r/\rho_0 (\rho_r - \rho_0)/(3\rho_0),
\end{equation}
by using $E_{\rm{sym}}(\rho) \approx E_{\rm{sym}}(\rho_0) + L \chi + \frac{1}{2!}K_{\rm{sym}}\chi^2$, which is a very good approximation to $E_{\rm{sym}}(\rho)$ for density less than about $2\rho_0$~\cite{ChenLW09,ChenLW11}. Taking $\rho_c = 0.11~{\rm fm}^{-3} \approx 2/3\rho_0$, one can obtain the following relations
\begin{eqnarray}\label{Eq:Lc-Esym}
L &  \approx & 3L_c/2 + K_{\rm{sym}}/9, \\
E_{\rm{sym}}(\rho_0) &  \approx &  E_{\rm{sym}}(\rho_c) +L_c/6 + K_{\rm{sym}}/162, \\
E_{\rm{sym}}(2\rho_0) &  \approx &  E_{\rm{sym}}(\rho_c) +2L_c/3 + 8K_{\rm{sym}}/81,
\end{eqnarray}
which indicate the $L$, $E_{\rm sym}(\rho_0)$ and $E_{\rm sym}(2\rho_0)$ are all linearly correlated with $L_c$ (and thus $\Delta r^{208}_{\rm{np}}$) for fixed $E_{\rm{sym}}(\rho_c)$ and small disturbance from $K_{\rm{sym}}$.
From the strong linear correlations
shown in Fig.~\ref{Fig_EsymNSkin}, one obtains $L = 83.1 \pm 24.7$~MeV, $E_{\rm{sym}}({\rho_0 }) = 34.3 \pm 1.7$~MeV, and $E_{\rm{sym}}({2\rho_0 }) = 62.8 \pm 15.9$~MeV. These results suggest a rather stiff symmetry energy around $\rho_0$, in contrast to the constraints $E_{\rm{sym}}({\rho_0 }) = 31.6 \pm 2.7$~MeV and $L = 58.9 \pm 16$~MeV~\cite{LiBA13,LiBA19EPJA}, or $E_{\rm{sym}}({\rho_0 }) = 31.7 \pm 3.2$~MeV and $L = 58.7 \pm 28.1$~MeV ~\cite{Oertel17RMP}, and $E_{\rm{sym}}({2\rho_0 }) = 47^{+23}_{-22}$~MeV~\cite{XieWJ20ApJ}, obtained by averaging essentially all the existing constraints.
In addition, {\it ab initio} coupled-cluster calculations~\cite{Hagen16NaturePhys} predict a rather soft symmetry energy of $37.8 \le L \le 47.7$~MeV and $25.2 \le E_{\rm{sym}}({\rho_0 }) = 30.4$~MeV, which are significantly smaller than our present constraints.
It is interesting to mention that
the present constraints are in surprisingly good agreement with the earlier constraint $L = 88 \pm 22$~MeV~\cite{ChenLW05PRC} obtained from transport model analyses~\cite{ChenLW05PRL,LiBA05PRC} on the isospin diffusion data~\cite{Tsang04} in heavy-ion collisions, as well as the constraints $E_{\rm{sym}}({\rho_0 }) = 35.0 \pm 1.5$~MeV and $L = 85.5 \pm 15.5$~MeV obtained from the analyses of isovector skin and isobaric analog states~\cite{Dan17}.

\begin{figure}[tbp]
\centering
\includegraphics[width=1.0\linewidth]{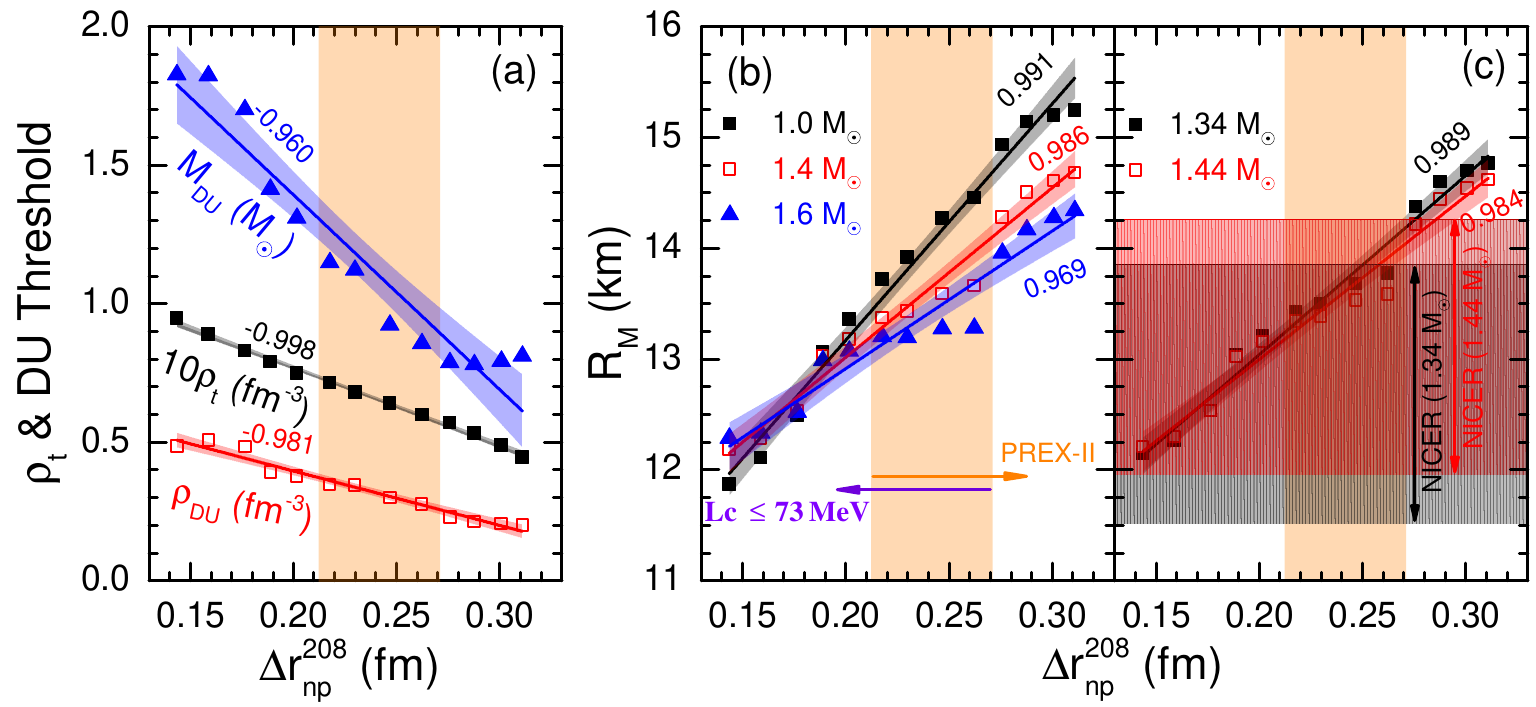}
\caption{Same as Fig.~\ref{Fig_EsymNSkin} but for the correlation with $\rho_t$, $\rho_{\rm DU}$ and $M_{\rm DU}$~(a), the radius $R_M$ of NS with mass $M = 1.0M_\odot$, $1.4M_\odot$ and $1.6M_\odot$~(b), and the $R_M$ with $M = 1.34M_\odot$ and $1.44M_\odot$~(c). The NICER constraints~\cite{Ril19NICER,Mil19NICER} are also included in panel~(c) for comparison.}
\label{Fig_NStarNSkin}
\end{figure}

Figure~\ref{Fig_NStarNSkin} shows the correlation of $\Delta r^{208}_{\rm{np}}$ with the crust-core transition density $\rho_t$, the threshold density $\rho_{\rm DU}$ and threshold NS mass $M_{\rm DU}$ above which the direct Urca~(DU) process $(n\rightarrow p+e^{-}+\bar{\nu}_{e},~p+e^{-}\rightarrow n+\nu_{e})$~\cite{Lat91} becomes possible, the radius $R_M$ of NS with mass $M = 1.0M_\odot$, $1.4M_\odot$ and $1.6M_\odot$ as well as $M = 1.34M_\odot$ and $1.44M_\odot$.
One sees the $\Delta r^{208}_{\rm{np}}$ exhibits a strong linear (anti-)correlation with all these NS properties, which together with the constraint $0.212~{\rm fm} \le \Delta r^{208}_{\rm{np}} \le 0.271~{\rm fm}$ allow us to obtain following information:
$\rho_t = 0.065\pm 0.010$~fm$^{-3}$,
$\rho_{\rm DU} = 0.313\pm 0.096$~fm$^{-3}$,
$M_{\rm DU} = (1.09 \pm 0.41)M_\odot$,
$R_{1.0} = 14.07\pm 0.91$~km,
$R_{1.4} = 13.66\pm 0.71$~km,
$R_{1.6} = 13.44\pm 0.68$~km,
$R_{1.34} = 13.72\pm 0.72$~km, and
$R_{1.44} = 13.62\pm 0.70$~km.
Our results suggest a relatively small $\rho_t$, implying the NS crust will have a small thickness, fractional mass, and moment of inertia~\cite{XuJ09ApJ}.
The $M_{\rm DU} = (1.09 \pm 0.41)M_\odot$
and $\rho_{\rm DU} = 0.313\pm 0.096$~fm$^{-3}$
indicate that the DU process will clearly occur in NSs with mass larger than $1.50M_\odot$ (central density larger than $0.409$~fm$^{-3}$).
Furthermore,
if $\Delta r^{208}_{\rm{np}}$ were larger than $0.25~{\rm fm}$, one obtains $M_{\rm DU} \lesssim 1.0M_\odot$, and this means the DU process will occur in essentially all the observed NSs. The DU process will enhance the emission of neutrinos and make it a more important process in the cooling of a NS~\cite{Lat91}.
This observation is particularly interesting given the fact that a fast neutrino-cooling process has been suggested by the detected x-ray spectrum of the NS in the low-mass x-ray binary MXB 1659-29~\cite{Bro18PRL}.
Nevertheless, it should be mentioned
that the NS cooling can be significantly influenced by nucleon pairing~\cite{Potekhin19AA,Page09ApJ}.

As for the NS radii,
very strong limits with a precision of $5\%$ have been obtained. In particular, our present results $R_{1.34} = 13.72\pm 0.72$~km and
$R_{1.44} = 13.62\pm 0.70$~km are in agreement with the NICER constraints~\cite{Ril19NICER,Mil19NICER} but have much better precision.
It is interesting to point out that
the NICER constraints have not been imposed in constructing the parameter sets of $L_c = 30 \sim 90$~MeV, implying in eSHF, they are compatible with $\Lambda_{1.4} \le 580$ and $M_{\rm{max}} \ge 2.05M_\odot$ as well as the flow data constraints on SNM EOS.
It should be noted that the NS radius depends on the poorly-known inner crust EOS~\cite{XuJ09ApJ,For16} but $\Lambda_{1.4}$ does not~\cite{ZhouY19PRD}, and thus it is relatively safer to use $\Lambda_{1.4}$ as a constraint.

\begin{figure}[tbp]
\centering
\includegraphics[width=0.9\linewidth]{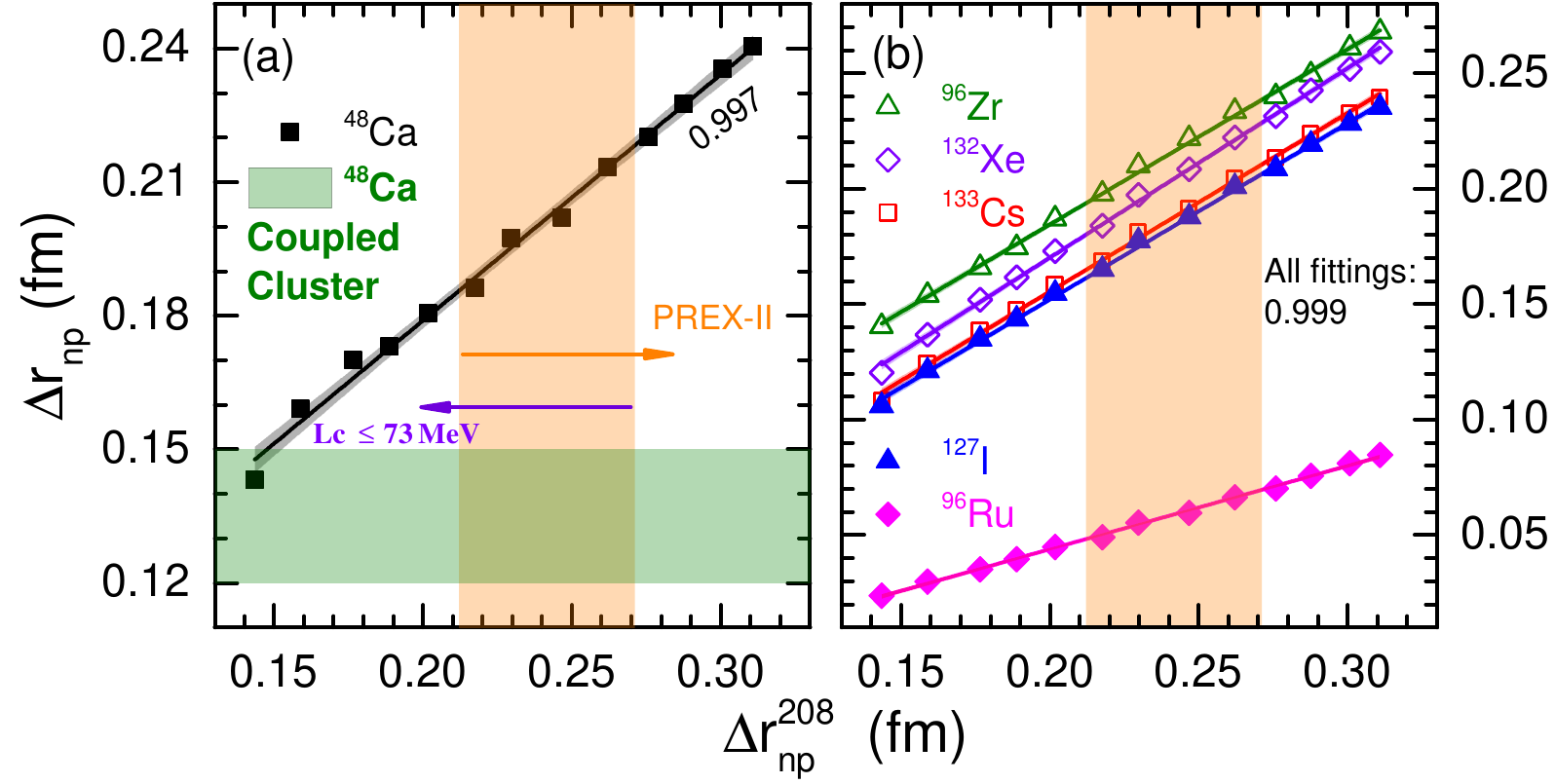}
\caption{Same as Fig.~\ref{Fig_EsymNSkin} but for the correlation with $\Delta r_{\rm{np}}$ of $^{48}$Ca~(a) and $^{96}$Zr, $^{96}$Ru, $^{127}$I, $^{133}$Cs, $^{132}$Xe~(b). The $\Delta r_{\rm{np}}$ of $^{48}$Ca from {\it ab initio} coupled-cluster calculations~\cite{Hagen16NaturePhys} is also included in panel~(a) for comparison.}
\label{Fig_NSkin}
\end{figure}

Shown in Fig.~\ref{Fig_NSkin} is the $\Delta r^{208}_{\rm{np}}$ vs the $\Delta r_{\rm{np}}$ of $^{48}$Ca, $^{96}$Zr, $^{96}$Ru, $^{127}$I, $^{133}$Cs, $^{132}$Xe, and very strong positive linear correlations are seen between the $\Delta r_{\rm{np}}$ of these nuclei.
Using $0.212~{\rm fm} \le \Delta r^{208}_{\rm{np}} \le 0.271~{\rm fm}$ together with the linear correlations, we obtain
$\Delta r^{48}_{\rm{np}}(\rm{Ca}) = 0.202\pm0.020$~fm for $^{48}$Ca,
$\Delta r^{96}_{\rm{np}}(\rm{Zr}) = 0.216\pm0.025$~fm for $^{96}$Zr,
$\Delta r^{96}_{\rm{np}}(\rm{Ru}) = 0.059\pm0.012$~fm for $^{96}$Ru,
$\Delta r^{127}_{\rm{np}}(\rm{I}) = 0.184\pm0.025$~fm for $^{127}$I,
$\Delta r^{133}_{\rm{np}}(\rm{Cs}) = 0.187\pm0.025$~fm for $^{133}$Cs, and
$\Delta r^{132}_{\rm{np}}(\rm{Xe}) = 0.204\pm0.027$~fm for $^{132}$Xe.

Particularly interesting is the $\Delta r^{48}_{\rm{np}}(\rm{Ca})$ as it has been predicted to be $0.12 \le \Delta r^{48}_{\rm{np}}(\rm{Ca}) \le 0.15$~fm from {\it ab initio} coupled-cluster calculations~\cite{Hagen16NaturePhys}, which is significantly smaller than our result $\Delta r^{48}_{\rm{np}}(\rm{Ca}) = 0.202\pm0.020$~fm.
At this point, we must mention that the Calcium Radius EXperiment~(CREX)~\cite{PREX-II} is expected to finish the data analysis on $\Delta r^{48}_{\rm{np}}(\rm{Ca})$ soon with a precision of $0.5\%$ (or $\pm0.02$~fm) for its $r_n$. CREX can thus provide a unique
bridge between {\it ab initio} approaches and density functional theory~(DFT). This is particularly important as the DFT (e.g., eSHF) is still the only realistic framework to investigate the physics of heavy nuclei and NSs.

The $\Delta r^{96}_{\rm{np}}(\rm{Zr})$ and $\Delta r^{96}_{\rm{np}}(\rm{Ru})$ are also very interesting since a recent study~\cite{LiHL20PRL} has demonstrated that the isobaric $^{96}$Zr+$^{96}$Zr and $^{96}$Ru+$^{96}$Ru collisions at relativistic energies can be used to extract the $\Delta r_{\rm{np}}$ of $^{96}$Zr and $^{96}$Ru with a weak model-dependence. The $\Delta r^{96}_{\rm{np}}(\rm{Zr})$ and $\Delta r^{96}_{\rm{np}}(\rm{Ru})$ are also crucial for the chiral magnetic effect search in isobaric collisions~\cite{XuHJ18PRL}. Our present results of $\Delta r^{96}_{\rm{np}}(\rm{Zr})$ and $\Delta r^{96}_{\rm{np}}(\rm{Ru})$ are particularly timely, because the data on these isobaric collisions at RHIC have been taken in 2018 and have been subject to a blinded analysis to assess the chiral magnetic effect.
In addition,
our results of $\Delta r^{127}_{\rm{np}}(\rm{I})$ and $\Delta r^{133}_{\rm{np}}(\rm{Cs})$ are critical for the information extraction of new physics~\cite{HuangXR19} via coherent elastic neutrino-nucleus scattering in the COHERENT experiment~\cite{Aki17}, while the $\Delta r^{132}_{\rm{np}}(\rm{Xe})$ is important for dark matter direct detection in liquid Xe detector~\cite{ZhengH14}.

Finally, it is instructive to see how our results change if the adopted constraints are varied. We note the upper limit of $L_c$ changes from $73$~MeV to $82$~MeV if the $\Lambda_{1.4} \le 580$~\cite{Abb18NSMerger} is altered into $\Lambda_{1.4} \le 720$ which seems to be favored if a NS maximum mass of $1.97M_\odot$ is imposed in the analysis of GW170817~\cite{GW170817MaxMass}.
In addition, varying the $M_{\rm{max}} \ge 2.05M_\odot$~\cite{Cro19Mmax} into the recently updated $M_{\rm{max}} \ge 2.01M_\odot$~\cite{Fon21Mmax} from PSR J0740+6620 only changes $L_c \le73$~MeV to $L_c \le74$~MeV, enhancing the pressure upper limit of SNM EOS constraint from the flow data by $10\%$ essentially does not change the limit $L_c \le73$~MeV, and
replacing $E_{\rm sym}(\rho_c) = 26.65$~MeV by $E_{\rm sym}(\rho_c) = 25.65(27.65)$~MeV leads to $L_c \le69(75)$~MeV.
We also note
replacing $\Lambda_{1.4} \le 580$ by $\Lambda_{1.4} \le 720$ leads to $0.212 \le \Delta r^{208}_{\rm{np}} \le 0.288$~fm, $E_{\rm{sym}}({\rho_0 }) = 34.8 \pm 2.1$~MeV, $L({\rho_0 }) = 89.0 \pm 30.6$~MeV, and $E_{\rm{sym}}({2\rho_0 }) = 65.7 \pm 18.7$~MeV.
At last, it should be noted that the present results are based on the conventional NS model in a single unified framework without considering possible new degrees of freedom (hyperons, meson condensates, quark matter, and so on) and modified gravity.

\textit{Conclusion.}---
We have demonstrated the symmetry energy slope parameter $L_c$ cannot be larger than $73$~MeV under the condition of $\Lambda_{1.4} \le 580$, and this leads to an upper limit of $\Delta r^{208}_{\rm{np}} \le 0.271$~fm.
This limit together
with the recent model-independent measurement on $\Delta r^{208}_{\rm{np}}$ from PREX-II leads to a rather large but very precise constraint of $0.212 \le \Delta r^{208}_{\rm{np}} \le 0.271$~fm, which suggests a rather stiff symmetry energy around $\rho_0$ and has critical implications on many issues in nuclear physics and astrophysics.
In particular, our present constraints on the symmetry energy and the neutron skin of $^{48}$Ca reveal serious tension with the predictions from {\it ab initio} coupled-cluster theory, and the soon coming data from CREX thus become extremely important.

\textit{Acknowledgments.}---
This work was supported by National SKA Program of China No. 2020SKA0120300 and the National Natural Science Foundation of China under Grant No. 11625521 and No. 11905302.

\end{document}